\documentclass[aps,prl, twocolumn,notitlepage, showkeys,showpacs,superscriptaddress,longbibliography]{revtex4-1}
\usepackage{epsf}
\usepackage{bm}
\usepackage{times}
\usepackage{amsfonts}
\usepackage{amssymb}
\usepackage{amsmath,mathtools}
\usepackage{array}
\usepackage{enumerate,dsfont,here,url}
\usepackage{dcolumn,multirow}
\usepackage{tikz-cd}
\usepackage{bbding}
\usepackage{tabularx, bbold}    
\usepackage[colorlinks=true,citecolor=blue,linkcolor=blue]{hyperref} 
\usepackage[normalem]{ulem}

\newcommand{\bk}{\bm{k}}

\newcommand{\mZ}{\mathbb{Z}}

\newcommand{\add}[1]{{\color{black} #1}}

\begin{document}

\title{High-throughput Investigations of Topological and Nodal Superconductors}
\author{Feng Tang}
\thanks{These two authors contribute equally to this work.}
\affiliation{National Laboratory of Solid State Microstructures and School of Physics, Nanjing University, Nanjing 210093, China and Collaborative Innovation Center of Advanced Microstructures, Nanjing University, Nanjing 210093, China}
\author{Seishiro Ono}
\thanks{These two authors contribute equally to this work.}
\affiliation{Department of Applied Physics, University of Tokyo, Tokyo 113-8656, Japan}
\author{Xiangang Wan}\email{xgwan@nju.edu.cn}
\affiliation{National Laboratory of Solid State Microstructures and School of Physics, Nanjing University, Nanjing 210093, China and Collaborative Innovation Center of Advanced Microstructures, Nanjing University, Nanjing 210093, China}
\author{Haruki Watanabe}\email{hwatanabe@g.ecc.u-tokyo.ac.jp}
\affiliation{Department of Applied Physics, University of Tokyo, Tokyo 113-8656, Japan}

\begin{abstract}
	 The theory of symmetry indicators has enabled database searches for topological materials in normal conducting phases, which has led to several encyclopedic topological material databases.  
	To date, such a database for topological superconductors is yet to be achieved because of the lack of information about pairing symmetries of realistic materials. In this work, sidestepping this issue, we tackle an alternative problem: the predictions of topological and nodal superconductivity in materials for each single-valued representation of point groups. Based on recently developed symmetry indicators for superconductors, we provide comprehensive mappings from pairing symmetries to topological or nodal superconducting nature for nonmagnetic materials listed in Inorganic Crystal Structure Database. 
	We quantitatively show that around 90\% of computed materials are topological or nodal superconductors when a pairing that belongs to a one-dimensional nontrivial irrep of point groups is assumed. When materials are representation-enforced nodal superconductors, positions and shapes of the nodes are also identified. These data are aggregated at \textit{Database of Topological and Nodal Supercoductors} (\url{https://tnsc.nju.edu.cn}). We also provide a subroutine \textit{Topological Supercon} (\url{http://toposupercon.t.u-tokyo.ac.jp/tms}), which allows users to examine the topological nature in the superconducting phase of any material themselves by uploading the result of first-principles calculations as an input.
	Our database and subroutine, when combined with experiments, will help us understand the pairing mechanism and facilitate realizations of the long-sought Majorana fermions promising for topological quantum computations.
\end{abstract}
\maketitle
Ever since heavy-fermion superconductors were discovered, intensive research on unconventional superconductors has been conducted, revealing several candidate materials~\cite{CeCu2Si2,UPt3,Sr2RuO4,UTe2,captas-sc}. While the nature of superconducting pairing of materials realized in actual experiments is often controversial, bulk nodes and surface states in the Bogoliubov quasi-particle spectrum are some of the few important experimental signatures of pairing symmetries of Cooper pairs. For example, 
a power-law behavior of the specific heat at low temperatures indicates the existence of nodes in the Bogoliubov quasi-particle spectrum.
The appearance of gapless surface states implies that the  superconductor is topological. 
In particular, Majorana surface states could be leveraged for topological quantum computations~\cite{Franz-RMP, Alicea-review, Wilczek}. 
Therefore, exploring nodal superconductors (NSCs) and topological superconductors (TSCs) is one of the necessary steps to elucidate the mechanism of unconventional superconductivity and to establish devices utilizing the novel property of TSCs. 

The past two decades have witnessed rapid developments in the theoretical understanding of topological phases of matter~\cite{RMP-Qi,RMP-Chiu}. 
The combinations of crystalline and internal symmetries give rise to a large variety of gapped and gapless topological phases, such as higher-order topological insulators~\cite{HOTI-S, HOTI-SA, HOTI-PRL-TSC, HOTI-Song}, Weyl/Dirac semimetals~\cite{Weyl,SMYoung, Na3Bi, Cd3As2}, and nodal-line semimetals~\cite{Balents-NL,Fang-NL,nodal-chain}. The recently established method of symmetry indicators (SIs)~\cite{Po-SI} and topological quantum chemistry~\cite{TQC} enables efficient diagnosis of  topological phases by examining irreducible representations (irreps) of space groups. These theories underlie recent large-scale topological material discoveries based on first-principles calculations~\cite{N-1, N-2, N-3, N-4}, which have indeed assisted experimentalists to find new topological materials~\cite{NM-BiBr}.  Very recently, the method of SIs has been extended to superconductors~\cite{SCSI-2, SCSI-3, SCSI-4, SCSI-5, SCSI-6, SCSI-7,SCSI-8}.

The fundamental distinction between the normal conducting phases and superconducting phases originates from the particle-hole symmetry (PHS), which naturally arises in the mean-field Bogoliubov-de Gennes Hamiltonian for superconductors.  The PHS gives rise to exotic phases, such as Bogoliubov Fermi surfaces~\cite{NSC-Surface-PRL} and Majorana corner modes~\cite{KitaevChain,HOTI-PRL-TSC,HOTSC-PRB-Khalaf}, with no counterparts in normal conducting phases. 
Unfortunately, the number of experimentally confirmed materials of NSCs and TSCs is much less than that of topological insulators and semimetals. 
Recalling the success of Refs.~\cite{N-1,N-2,N-3,N-4}, it is tempting to think that constructing such a database for superconductors would be a catalyst for discovering new candidates of topological superconductors. However, this task is quite challenging because pairing symmetries for the majority of materials are not known.

In this work, we sidestep this difficulty and take an alternative approach: We apply the newly developed SIs for superconductors to materials listed in a database,  assuming that the materials become superconductors at low temperatures, and establish a comprehensive database summarizing the topological and nodal nature of materials in superconducting phases.
We should emphasize that it is not our ambition to predict topological and nodal superconductivity for an accurate superconducting pairing.
Even when the actual pairing symmetry, of which identification usually requires cautious studies, is not known, computing SIs for each possible pairing symmetry is still helpful in elucidating the pairing mechanism of unconventional superconductivity. These results could be compared with experimental results and reduce the candidates of pairing symmetries. We summarize our result of diagnosis of topological and nodal nature for all nonmagnetic materials in superconducting phases in \href{https://tnsc.nju.edu.cn}{\textit{Database of Topological and Nodal Supercoductors}}.
We also provide a subroutine \href{http://toposupercon.t.u-tokyo.ac.jp/tms}{\textit{Topological Supercon}}, which can diagnose TSC/NSC in any material synthesized in the future.  
\add{This program aims to quickly diagnose possible TSC/NSC in various conditions that are not covered in our database, for example, the cases where the SOC is completely absent.}

\begin{figure}[t]
	\includegraphics[width=1\columnwidth]{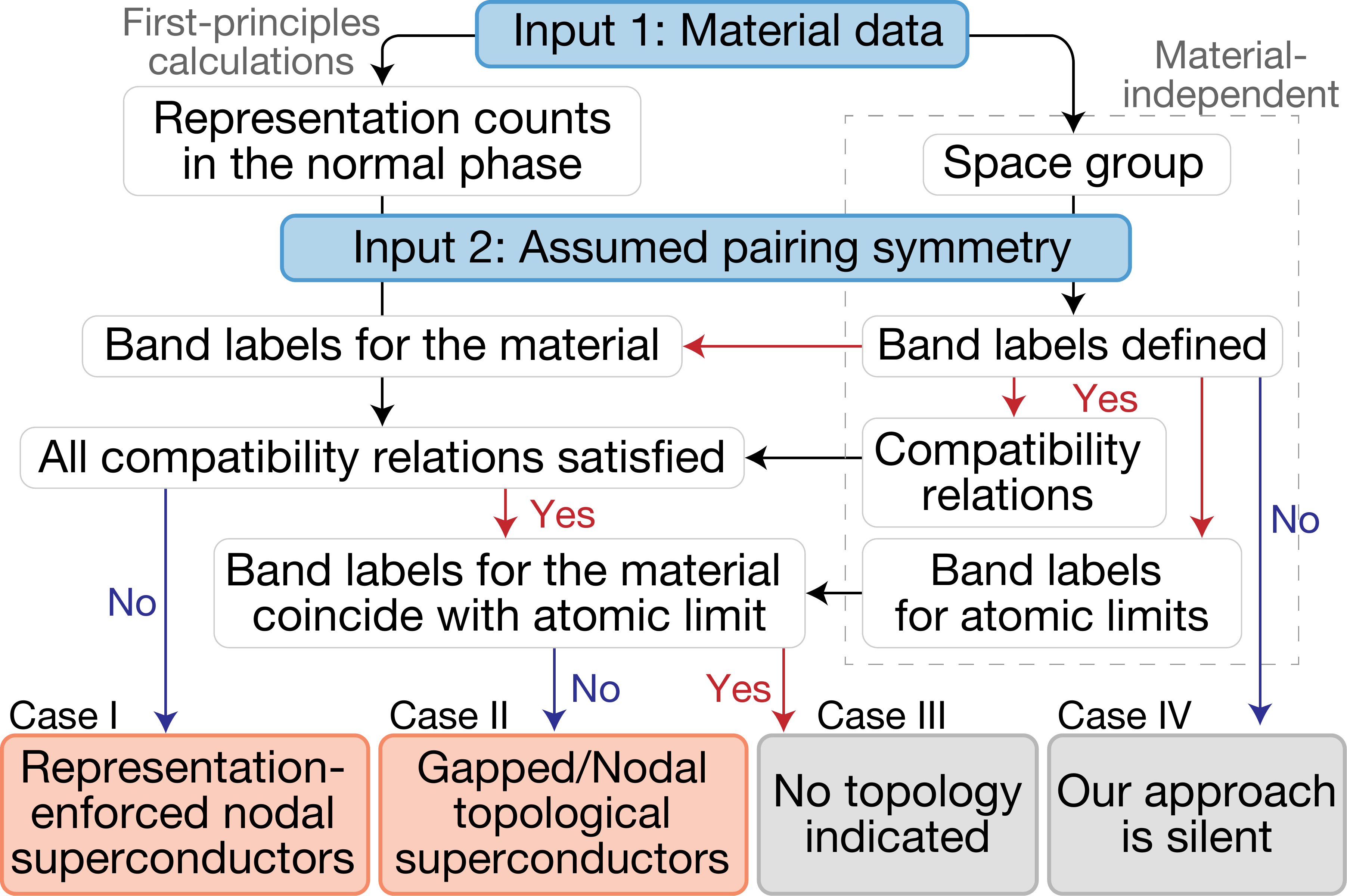}\\
	\caption{\label{flow}\textbf{Flow chart of our diagnostic scheme.} 
		For a given material, we identify the space group and irreducible representations of the normal phase. Assuming a pairing symmetry, we determine whether band labels are defined from the space group. If not, our approach is silent. Otherwise, we obtain compatibility relations and band labels of atomic limits. Using them, we diagnose the topological and nodal superconducting nature of the material.
	}
\end{figure}
{\it Diagnostic scheme---}Here we give a brief summary of our diagnostic scheme illustrated in Fig.~\ref{flow} (See Supplemental Materials (SM) for more details).
In general, the full diagnosis of the topological properties of a material requires (i) the complete information on its wave function over the entire Brillouin zone and (ii) the complete list of topological indices to be computed. However, these are often inaccessible for realistic materials, especially for superconductors.
Space group representations  of the wave function at high-symmetry momenta  enable us to efficiently diagnose topological and nodal nature of the system~\cite{Po-SI,SCSI-5, SCSI-6}.  Based on this idea,  
we focus only on zero-dimensional topological invariants of systems, which we call band labels (BLs) in this work. BLs are defined for each irrep at each high-symmetry momentum and they altogether form a list  $\bm{n}_{\text{SC}}$.

To realize a gapped Bogoliubov quasi-particle spectrum, BLs in $\bm{n}_{\text{SC}}$ are not fully independent and are subject to symmetry constraints called compatibility relations (CRs). If some CRs are violated, the superconductor is a representation-enforced NSC. By inspecting the violations of CRs, we can determine the positions and shapes of nodes such as point, line, and surface~\cite{SCSI-7}.
When all CRs are satisfied, the superconductor may be fully gapped or may still exhibit nodes of a topological origin.  In this case, we compare the set of BLs $\bm{n}_{\text{SC}}$ with those of atomic limit superconductors  that are constructed from localized orbitals at Wyckoff positions and thus are topologically trivial. Since BLs are topological invariants, a superconductor whose $\bm{n}_{\text{SC}}$ is different from any atomic limit must be topologically nontrivial, i.e., the superconductor is a TSC or a topological NSC. 

For realistic materials, \add{once we assume a pairing symmetry,} we can determine $\bm{n}_{\text{SC}}$ from the first-principles calculations based on the weak pairing assumption: the energy scale of the Cooper pair potential is much smaller than that of normal conducting phases~\cite{Sato-Ando}. In such a case, BLs for the superconducting phase can be deduced from the numbers of occupied bands characterized by irreps in the normal conducting phase. Using the obtained BLs, we analyze topological and nodal nature as explained in the preceding discussions. After performing the analyses, each material for each pairing falls into one of the following cases: Case I, representation-enforced NSC; Case II, symmetry-diagnosable TSC or topological NSC; Case III, topologically trivial or not symmetry-diagnosable TSC; Case IV, the case where no BLs can be defined at any high-symmetry momenta. In particular, as far as the time-reversal symmetry (TRS) is assumed in the normal conducting phase, Case IV occurs if and only if the trivial pairing symmetry is assumed.

\add{To precisely formulate the pairing symmetry of the superconducting order parameter, 
suppose that $u_{\bk}(g)$ is the unitary representation of a space group symmetry $g$ in the normal phase. We introduce $m$ independent basis matrices $\Delta_{i}(\bk)$ that transform under $g$ as $u_{\bk}(g)\Delta_{i}(\bk)u_{-\bk}^{T}(g) = \sum_{j=1}^m\Delta_{j}(g\bk)[\mathcal{D}(g)]_{ji}$ and expand $\Delta(\bk)$ as $\sum_{i = 1}^{m}\eta_i \Delta_{i}(\bk)$.  Here $\mathcal{D}(g)$ is a $m$-dimensional single-valued representation of the point group.
We say the pairing is nontrivial when $\mathcal{D}(g)$ is a nontrivial irrep. 
When $\mathcal{D}(g)$ is multi-dimensional, 
the space group of the superconducting phase is lowered to its subgroup for which $(\eta_1,\eta_2,\cdots,\eta_m)$ is a simultaneous eigenvector of $\mathcal{D}(g)$.
We assume that the lowered group is a maximal subgroup of the original space group, as in most cases of materials.
The irreps of occupied bands in the lowered symmetry setting can be deduced from the irreps of the original group (see SM-I and II for more details). 
}



\add{It should be emphasized that we do not need to know which orbitals form the pair in this framework.
	In particular, our symmetry-based analysis cares only about the symmetry property of the pairing and does not distinguish between multi-orbital pairings and single-orbital pairings. This makes the analysis more accessible and systematic.}

{\it Material investigation---}We conduct calculations for 27,208 out of more than 180,000 materials listed in Inorganic Crystal Structure Database (ICSD)~\cite{icsd}, excluding  nonstoichiometric materials and those containing typical magnetic ions (see SM). We find 10,822 of them have some sort of Fermi surfaces, and in this section we focus on them. For each nontrivial pairing symmetry, we determine the topological property of the superconductor. 

We find that the majority of the results fall into Case I or II. For example, when restricting to nontrivial one-dimensional irreps, 88.9\% of computed materials always fall into Case I or II, regardless of the choices of nontrivial one-dimensional irreps. 
This is consistent with the empirical rule that a superconductor is likely to be nontrivial when \add{a nontrivial} pairing is assumed.
Complete lists of computed materials and the percentage for each space group are included in SM-III.
It is worth noting that 22 space groups have high percentages, which might guide the search for TSCs or NSCs. For example, once \add{nontrivial} pairings are formed, 100\% of computed materials in space group No.~191 are TSCs or NSCs. Therefore, one can expect better chances of discovering new TSCs and NSCs by exploring the materials in these space groups.

We should point out that our database contains many materials whose superconductivity has not been confirmed yet. However, most of them could be superconductors at extremely low temperatures. Indeed, superconductivity in YbRh$_2$Si$_2$ has recently been observed at around eight milliKelvin~\cite{YbRh2Si2}.

We also note that our analysis is sensitive to the Fermi level determined by the first-principles calculation since BLs are obtained from the counts of irreps of occupied bands below the Fermi energy.  Furthermore, even when the undoped material does not exhibit superconductivity as is the case in large-gap insulators, doped one sometimes does. For these reasons, we also perform calculations for several different Fermi levels for each of 27,208 materials (see SM-III for the details).



\begin{figure}[t]
	\begin{center}
		\includegraphics[width=0.9\columnwidth]{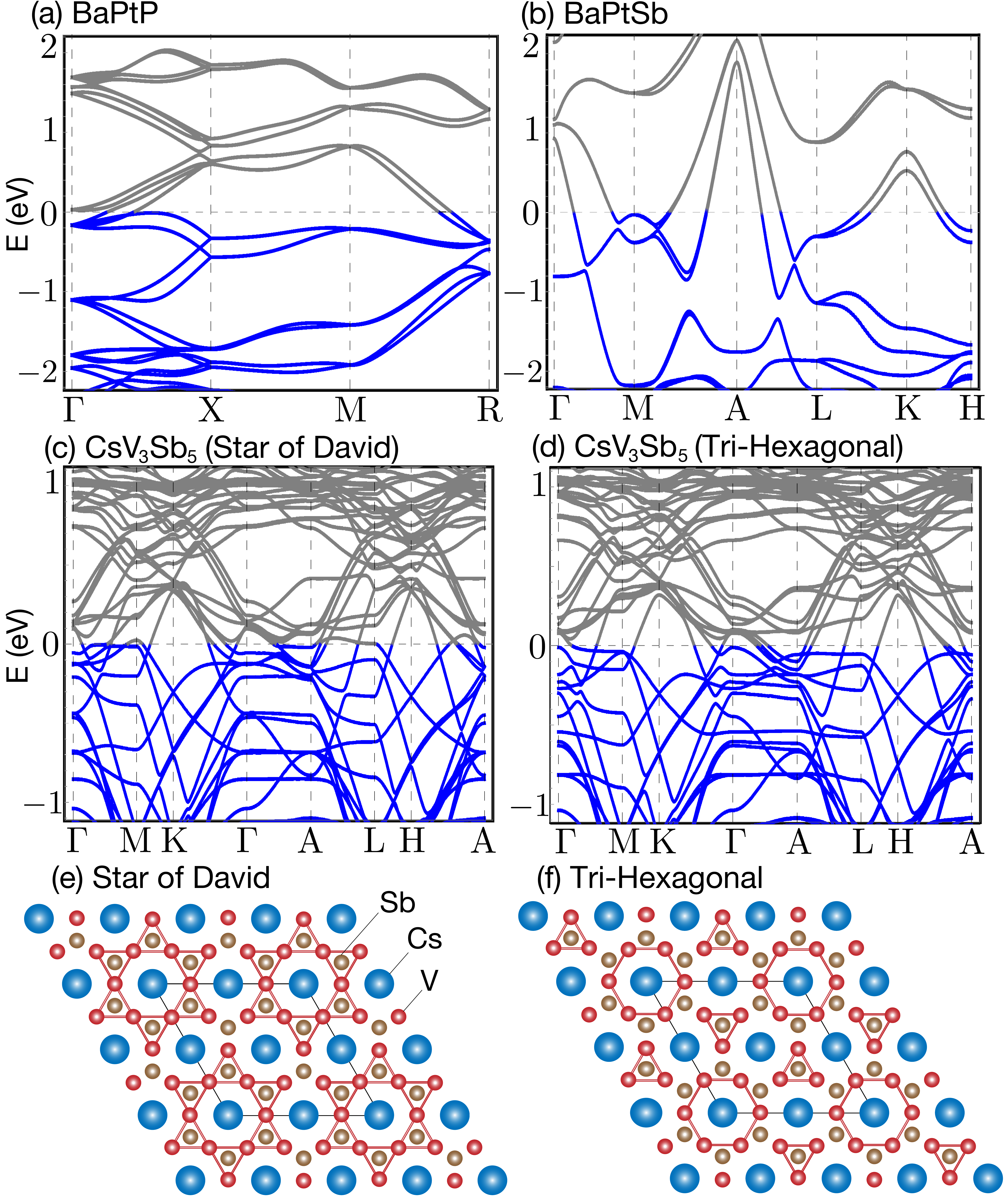}
		\caption{\textbf{Structures and band structures of material examples.}
		(a-d) Band structures of BaPtP (a); BaPtSb (b); CsV$_3$Sb$_5$ in `Star of David' (c) and `Tri-Hexagonal' (d) structures. (e, f) `Star of David' (e) and `Tri-Hexagonal' (f) crystal structures of CsV$_3$Sb$_5$ introduced in Ref.~\onlinecite{AV3Sb5_Fermi}.}
		\label{fig3}
	\end{center}
\end{figure}
{\it Material Examples in the database---}Here, we pick two materials BaPtP and BaPtSb from our database as examples.
Superconductivity was recently observed in these compounds~\cite{baptp-sc,baptsb-sc}.
Interestingly, their constituent elements belong to the same families as those in similar compounds CaPtAs and SrPtAs, which are candidates of uncovenctional superconductors~\cite{captas-sc, srptas-sc}. 
The space groups of these materials, $P2_13$ (No.~198)  for BaPtP and $P\bar{6}m2$ (No.~187) for BaPtSb, lack inversion symmetry~\cite{icsd}. In fact, one of the advantages of our method is that it does not require inversion symmetry.
The electronic band structures obtained by first-principles calculations are shown in Fig.~\ref{fig3} (a) and (b).
The results of our symmetry-based analysis are tabulated in Table~\ref{results}.  

Let us start with BaPtP, which is predicted to host unconventional excitations with four- and six-fold degeneracies~\cite{RhSi,N-1}. The point group  $T$ has three 1D single-valued representations: $A$, $^{1}E$, and $^{2}E$. The latter two break TRS.
We find that a superconductor with $^{1}E$ or $^{2}E$ pairing is a representation-enforced NSC, in which CRs along the $\Gamma$-R line are violated. 
The resulting nodes along this line turn out to be part of a surface node.


\add{On the other hand, the point group has a three-dimensional irrep $T$, which can be decomposed into $B_1\oplus B_2\oplus B_3$ of point group $D_2$ or $A\oplus ^1E \oplus ^2E$ of point group $C_3$. 
Here we assume that $\mathcal{D}(g)$ is either one of $B_1$, $B_2$, or $B_3$ of $D_2$.
Then the resulting space group is $P2_12_12_1$ (No.~19). 
We obtain irreps of occupied bands in $P2_12_12_1$ from those in $P2_13$ and perform the same analyses.  
The superconducting phase is predicted to have nodal lines.}

Next, we discuss BaPtSb. The point group is $D_{3h}$ containing four 1D single-valued representations: $A''_2, A'_2,A''_1$ and $A'_1$, all of which preserve TRS.  
\add{There are also two-dimensional irreps but here we focus on 1D irreps.}
We find that a superconductor with $A''_2$ or $A'_2$ pairing is a representation-enforced NSC. CRs along various lines are violated, as shown in Table~\ref{results}. Since all of the lines are in $\Gamma$-A-H-K or M-L-H-K planes, the violations result in line nodes in these planes~\cite{SCSI-7}. In contrast, all CRs are satisfied for the $A''_1$ pairing. Thus we diagnose the topology by SIs. We find that the material belongs to the entry $(0,1,0,0,5,5) \in (\mZ_2)^4 \times (\mZ_6)^2$ of SIs, which indicates that  the mirror Chern number on the $k_z=\pi$ equals $1$ mod $3$ while that on the $k_z=0$ plane is $0$ mod $3$. 

Finally, let us explain how to utilize our database for studies of unconventional superconductivity in realistic materials.
There is a fundamental distinction between our database and those for normal states~\cite{N-1,N-2,N-3,N-4}. 
The SI method for superconductors~\cite{SCSI-6,SCSI-7} requires the pairing symmetry of the material as input.
However, it is quite challenging to determine pairing symmetry. It requires careful experimental studies and the result is sometimes controversial. 

Furthermore, the trivial pairing symmetry (the $A$ representation for BaPtP and the $A'_1$ representation for BaPtSb in the above examples) in the presence of TRS always falls into Case IV, in which our symmetry-based analysis has no predicting power. Thus, one cannot exclusively conclude that the superconducting phase of a given material is topological or nodal just by referring to our database.
Although our database seems to be useless at first glance, we argue that this is not the case.
While our method cannot determine pairing symmetries themselves, the comprehensive correspondences between pairing symmetries and topological and nodal superconducting nature for materials in ICSD enable us to go back and forth between experiments and our database.
For example, if one obtains an experimental result such as the specific heat and the magnetic penetration depth of the material, then one should refer to the database and compare the results with nodal and topological nature listed in the database. This way one can reduce the possible pairing symmetries of the material consistent with the experiment.

\begin{table}
	\begin{center}
		\caption{\label{results}
			\textbf{Results of symmetry-based diagnosis for BaPtP, BaPtSb, and CsV$_3$Sb$_5$.}
			The first column represents material names.
			The second one is the point group of the superconducting phase, and symbols in the third column mean assumed irreducible representations of the pairing. Definitions of these representations are included in Supplemental Materials. 
			The fourth column indicates Case I--IV for each pairing. Symbols [S], [L], and [P] for Case I specify the shape of the nodes, surface, line, and point, respectively.
			The last column represents the paths where CRs are broken for Case I and  the entry of symmetry indicators for Case II. 
			For Case II of BaPtSb and CsV$_3$Sb$_5$, the entries are in $(\mZ_2)^4 \times (\mZ_6)^2$ and $ (\mZ_2)^4 \times \mZ_{12} \times \mZ_{24}$, respectively.
		}
		\begin{tabular}{c|c|c|c|c}
			\hline
			 Material& PG& Irrep $\mathcal{D}(g)$ & Case& Nodes and Topology \\
			\hline
			\multirow{6}{*}{\shortstack{BaPtP \\ (ICSD:\\ 059191)}}&\multirow{3}{*}{$\text{T}$}&$\text{}^2E$&I [S]&$\Gamma$-R\\
			&&\multirow{1}{*}{$\text{}^1E$}&\multirow{1}{*}{I [S]}&$\Gamma$-R\\
			&&$\text{A}$&IV&$-$\\
			& \multirow{3}{*}{$D_2$ }& $B_1$ & I [L] & $\Gamma$-Z\\
			&& $B_2$ & I [L] & $\Gamma$-Y\\
			&& $B_3$ & I [L] & $\Gamma$-X\\
			\hline
			\multirow{4}{*}{\shortstack{BaPtSb \\ (ICSD:\\ 059186)}}&\multirow{4}{*}{$D_{3h}$}&$A''_2$&I [L]&K-H\\
			&&\multirow{1}{*}{$A'_2$}&\multirow{1}{*}{I [L]}&$\Gamma$-K, M-K, K-H\\
			&&\multirow{1}{*}{$A''_1$}&\multirow{1}{*}{II}& $(0, 1, 0, 0, 5, 5)$\\
			&&$A'_1$&IV&$-$\\
			\hline
			\multirow{8}{*}{\shortstack{CsV$_3$Sb$_5$ \\ Star of David}}&\multirow{8}{*}{$D_{6h}$}&$B_{1u}$&I [P]&$\Gamma$-A\\
			&&\multirow{1}{*}{$B_{1g}$}&\multirow{1}{*}{I [L]}&$\Gamma$-A\\
			&&\multirow{1}{*}{$B_{2u}$}&\multirow{1}{*}{I [P]}&$\Gamma$-A\\
			&&\multirow{1}{*}{$B_{2g}$}&\multirow{1}{*}{I [L]}&$\Gamma$-A\\
			&&\multirow{1}{*}{$A_{2u}$}&\multirow{1}{*}{I [P]}&$\Gamma$-A\\
			&&\multirow{1}{*}{$A_{2g}$}&\multirow{1}{*}{I [L]}&$\Gamma$-A\\
			&&\multirow{1}{*}{$A_{1u}$}&\multirow{1}{*}{II}&$(0, 0, 0, 1, 8, 8)$\\
			&&$A_{1g}$&IV&$-$\\
			\hline
			\multirow{8}{*}{\shortstack{CsV$_3$Sb$_5$ \\ Tri-Hexagonal}}&\multirow{8}{*}{$D_{6h}$}&$B_{1u}$&I [P]&A-L\\
			&&\multirow{1}{*}{$B_{1g}$}&\multirow{1}{*}{I [L]}&$\Gamma$-K, M-K, A-L, L-H, $\Gamma$-A\\
			&&\multirow{1}{*}{$B_{2u}$}&\multirow{1}{*}{I [P]}&$\Gamma$-K, M-K, L-H\\
			&&\multirow{1}{*}{$B_{2g}$}&\multirow{1}{*}{I [L]}&$\Gamma$-K, M-K, A-L, L-H, $\Gamma$-A\\
			&&\multirow{1}{*}{$A_{2u}$}&\multirow{1}{*}{I [P]}&$\Gamma$-A\\
			&&\multirow{1}{*}{$A_{2g}$}&\multirow{1}{*}{I [L]}&$\Gamma$-K, M-K, A-L, L-H, $\Gamma$-A\\
			&&\multirow{1}{*}{$A_{1u}$}&\multirow{1}{*}{II}&$(0, 0, 1, 0, 9, 11)$\\
			&&$A_{1g}$&IV&$-$\\
			\hline
		\end{tabular}
	\end{center}
\end{table}

{\it Example of subroutine---}While we perform comprehensive computations for materials listed in ICSD, one may often encounter a situation where one wants to vary the composition of materials, to apply strains and pressures, or even to study materials not listed on the database.  To this end, we develop a subroutine that enables users to run an automated calculation of our diagnostic scheme by uploading a result of first-principles calculations in a particular format as the input. 
\add{
	Our subroutine is designed to be complementary to the database we developed, and can handle various situations not considered there. For example, one can choose different Altland-Zirnbauer classes~\cite{AZclass} that may lack the TRS. One can also examine multi-dimensional representations which break the space group further than maximal subgroups by preparing the input for the lowered symmetry setting.
}

To demonstrate the usage of the subroutine, here we discuss a new kagome metal CsV$_3$Sb$_5$, which has been recently synthesized~\cite{ACsV3Sb5_crystal,CsV3Sb_SC} and attracted much attention because of its various orders such as superconductivity and charge-density wave.
As examples, we consider two distorted crystal structures due to a charge-density wave order,  ``Star of David~\cite{AV3Sb_DFT,AV3Sb5_Fermi}'' and ``Tri-Hexagonal~\cite{AV3Sb5_Fermi}'' illustrated in Fig.~\ref{fig3} (g) and (h), both of which crystallize into space group $P6/mmm$ (No.~191), and assume that no further symmetry breaking occurs.

The outputs of our subroutine for these structures are summarized in Table~\ref{results} for one-dimensional $\mathcal{D}(g)$.
The predicted results are mostly representation-enforced NSCs unless the paring symmetry is $A_{1u}$ or $A_{1g}$.
The shapes of the nodes are point type for odd-parity cases and line types for even-parity cases.
By contrast, a fully gapped TSC may be realized for the $A_{1u}$ pairing.
As for the Star of David structure, the nontrivial SI implies that the mirror Chern number is $0$ mod $6$ on the $k_z=0$ plane and $2$ mod $6$ on the $k_z=\pi$ plane.  The three-dimensional winding number must be even (but could also be $0$).  On the other hand, for the Tri-Hexagonal structure, the mirror Chern number is $2$ and $1$ mod $6$ on both $k_z=0 $ and $\pi$ planes. The three-dimensional winding number must be odd, implying that the gapped TSC is a strong TSC.
The $A_{1g}$ paring is Case IV and our approach has no predicting power as mentioned before.
\add{
We include the results for multi-dimensional irreps in SM-III.}

Our results impose strong constraints on possible pairings in the gapped phase reported recently~\cite{CsV3Sb5_FSC}. 
That is, pairings belonging to 1D single-valued representations of the point group except for $A_{1u}$ and $A_{1g}$ must be excluded, because a full gap is prohibited for these parings.  
For this material, nodal superconductivity has also been observed~\cite{CsV3Sb5_NSC}. 
The relation between shapes of the nodes and the parity of the pairing, found above, will also help us identify the actual pairing symmetry of the material.

{\it Conclusions---}In summary, we applied the theory of SIs for superconductors to high-throughput identifications of  TSCs or NSCs in a nonmagnetic materials database.
We performed a comprehensive investigation and constructed  a database that lists topological and nodal nature of materials (see \href{https://tnsc.nju.edu.cn}{\textit{Database of Topological and Nodal Supercoductors}}). We also developed a subroutine \href{http://toposupercon.t.u-tokyo.ac.jp/tms}{\textit{Topological Supercon}}, which can be used by uploading a result of first-principles calculations.


\add{Topological flat bands near the Fermi energy open a promising avenue for unconventional superconductivity~\cite{flat-4,flat-3,flat-5,flat-1,flat-2}.
A recent comprehensive investigation of topological flat bands~\cite{flat-5} reveals that many materials in ICSD possess topological flat bands. 
Since materials in our database largely overlap with them, studying materials listed in these two databases is a promising direction for discovering new topological/nodal superconductors (see Table IV of SM-III).
}

Our database and subroutine can aid the determination of the pairing symmetry of superconductors. Results in this work immediately present implications of experimental signatures such as boundary states, specific heat properties, or other measurements.
An enormous number of materials predicted in this work are expected to facilitate future explorations of exotic states such as Majorana fermions, newly proposed response of TSCs~\cite{response_TSC}, and Bogoliubov Fermi surfaces.

\begin{acknowledgments}
S.O. thanks Hoi Chun Po and Ken Shiozaki for earlier collaborations on related topics. S.O. and H.W. thank Motoaki Hirayama for fruitful discussions. F.T. and X.W. thank Ge Yao for helps in computers and constructing website. F.T. thanks Kai Li, Zheng-Wei Zuo, and Xiaoyu Zhu for helpful discussions.
F.T. and X.W. were supported by the National Key R\&D Program of China (Grants No.~2017YFA0303203 and No.~2018YFA0305704), the National Natural Science Foundation of China (NSFC Grants No.~11834006, No.~51721001, and No.~11790311) and the excellent program at Nanjing University. X.W. also acknowledges the support from the Tencent Foundation through the XPLORER PRIZE.
S.O. is supported by The ANRI Fellowship and KAKENHI Grant No.~JP20J21692 from the Japan Society for the Promotion of Science. 
H.W. is supported by JST PRESTO grant NO.~JPMJPR18LA. 

F.T. and S.O. contributed equally to this work.
\end{acknowledgments}

\bibliography{sc}
\end{document}